\documentclass{Interspeech2024}

\interspeechcameraready

\title{One-pass Multiple Conformer and Foundation Speech Systems Compression and Quantization Using An All-in-one Neural Model}
\name[affiliation={1,*}]{Zhaoqing}{Li}
\name[affiliation={1,*}]{Haoning}{Xu}
\name[affiliation={1}]{Tianzi}{Wang}
\name[affiliation={2}]{Shoukang}{Hu}
\name[affiliation={1}]{Zengrui}{Jin}
\name[affiliation={1}]{Shujie}{Hu}
\name[affiliation={1}]{\\Jiajun}{Deng}
\name[affiliation={1}]{Mingyu}{Cui}
\name[affiliation={3}]{Mengzhe}{Geng}
\name[affiliation={1}]{Xunying}{Liu}

\address{
  $^1$The Chinese University of Hong Kong; $^2$Nanyang Technological University; \\$^3$National Research Council Candada\thanks{* Equal Contribution}}
\email{zqli@se.cuhk.edu.hk, hnxu@se.cuhk.edu.hk, xyliu@se.cuhk.edu.hk}

\keywords{ASR, model compression, model quantization, self-supervised learning, knowledge distillation, transformer}

\usepackage{multirow}
\usepackage{colortbl}  %彩色表格需要加载的宏包
\usepackage{xcolor}
\usepackage{array}
\usepackage{float}
\usepackage{cite}

\usepackage{bbding}
\usepackage{pifont}
\newcommand{\cmark}{\ding{51}}
\newcommand{\xmark}{\ding{55}}

\begin{document}
\bstctlcite{IEEEexample:BSTcontrol}

\maketitle
\begin{abstract}
    
    % Practical deployment of ASR systems on ad-hoc devices such as mobile phones and smartwatches require their varying computational resource constraints to be met. This paper proposes a novel one-pass multiple ASR systems joint compression and quantization approach using a single all-in-one neural model. A single system compression cycle allows a large number of target systems, for example, up to \red{8},  with different compression ratios that are achieved using varying Encoder depths, widths and quantization precision settings to be simultaneously constructed, without the need of compressing and training individual target systems one by one.  Experiments conducted the 300-hr Switchboard dataset trained Conformer and LibriSpeech-100hr fine-tuned wav2vec2.0 SSL models consistently demonstrate the multiple ASR systems being compressed in a single pass produced word error rate (WER) comparable to, or lower by up to \red{1.01}\% absolute (\red{6.98}\% relative) than, those obtained using individually compressed baseline systems of equal complexity.  An overall system compression and training time speed up ratio of \red{3.4x} was achieved.  Maximum model size compression ratio of \red{13.3x} and ?? times were obtained on the two tasks over the 32-bit full-precision baseline Conformer and wav2vec2.0 models respectively while incurring non-statistically significant or negligible WER increase.
    
    %Practical deployment of ASR systems on ad-hoc devices requires varying computational resource constraints to be met. 
    We propose a novel one-pass multiple ASR systems joint compression and quantization approach using an all-in-one neural model. A single compression cycle allows multiple nested systems with varying Encoder depths, widths, and quantization precision settings to be simultaneously constructed without the need to train and store individual target systems separately. Experiments consistently demonstrate the multiple ASR systems compressed in a single all-in-one model produced a word error rate (WER) comparable to, or lower by up to 1.01\% absolute (6.98\% relative) than individually trained systems of equal complexity. A 3.4x overall system compression and training time speed-up was achieved. Maximum model size compression ratios of 12.8x and 3.93x were obtained over the baseline Switchboard-300hr Conformer and LibriSpeech-100hr fine-tuned wav2vec2.0 models, respectively, incurring no statistically significant WER increase.
    
\end{abstract}

\section{Introduction}

Current automatic speech recognition (ASR) systems based on Transformer~\cite{vaswani2017attention} and their variants~\cite{gulati2020conformer,peng2022branchformer} are becoming increasingly complex for practical applications. Such trend is further exasperated by the emergence and rapid proliferation of self-supervised learning (SSL) based speech foundation models represented by wav2vec2.0~\cite{baevski2020wav2vec}, HuBERT~\cite{hsu2021hubert}, data2vec~\cite{baevski2022data2vec}, and WavLM~\cite{chen2022wavlm}. Their practical deployment on ad-hoc edge devices such as mobile phones, self-driving cars, and smartwatches require their respective computational resource constraints to be met. To this end, there is a pressing need to develop high-performance speech AI model compression techniques that flexibly account for the fine-grained performance-efficiency trade-off granularity across diverse user devices.

Prior research for Transformer based speech processing models has largely evolved into two categories: \textbf{1) architecture compression} methods that aim to minimize the Transformer model structural redundancy measured by their depth, width, sparsity, or their combinations using techniques such as pruning~\cite{pru3,pru5,jiang2023accurate}, low-rank matrix factorization~\cite{lr3,lilossless} and distillation~\cite{rathod2022multi,park2023conformer}; and \textbf{2) low-bit quantization} approaches that use either uniform~\cite{quant1,quant2,quant3,qiu2023role}, or mixed precision~\cite{lilossless,quant5} settings. A combination of both architecture compression and low-bit quantization approaches has also been studied to produce larger model compression ratios~\cite{lilossless}. Architecture compression techniques for the recently emerging SSL speech foundation models have also been developed using weight pruning~\cite{pru4,hj,lodagala2023pada}, distillation~\cite{distilhubert,dist1,dist2,dist3,chang2023colld,cho2023sd,de2023distilling,wang2022lighthubert}, or both~\cite{dphubert}. Only very limited prior research~\cite{deepcomp,quant4,usmlite} studied SSL speech model quantization. 

The above existing researches suffer from the following limitations: \textbf{1) weak scalability} when being used to produce compressed systems of varying target complexity that are tailored for diverse user devices. The commonly adopted approach requires each target compressed system with the desired size to be individually constructed, for example, in~\cite{park2023conformer,quant1,quant3} for Conformer models, and similarly for SSL foundation models such as DistilHuBERT~\cite{distilhubert}, FitHuBERT~\cite{dist1}, DPHuBERT~\cite{dphubert}, PARP~\cite{pru4}, and LightHuBERT~\cite{wang2022lighthubert} (no more than 3 systems of varying complexity were built). \textbf{2) limited scope of system complexity attributes} covering only a small subset of architecture hyper-parameters based on either network depth or width alone~\cite{pru3,pru5,lr3,once,uslim}, or both~\cite{jiang2023accurate,rathod2022multi,park2023conformer,hou2020dynabert}, while leaving out the task of low-bit quantization, or vice versa~\cite{quant1,quant2,quant3,qiu2023role,quant5,deepcomp,quant4,usmlite}. This is particularly the case with the recent \textbf{HuBERT} model distillation research~\cite{distilhubert,dist1,dist2,cho2023sd,de2023distilling,wang2022lighthubert,dphubert} that are focused on architectural compression alone. \textbf{3) restricted application} to either standard supervised-learning based Transformers trained on a limited quantity of data~\cite{pru3,pru5,jiang2023accurate,lr3,lilossless,rathod2022multi,park2023conformer,quant1,quant2,quant3,qiu2023role,quant5} or their SSL foundation models alone~\cite{pru4,hj,lodagala2023pada,distilhubert,dist1,dist2,dist3,chang2023colld,cho2023sd,de2023distilling,wang2022lighthubert,dphubert,deepcomp,quant4,usmlite}. Hence, a more comprehensive study on both model types is desired to offer further insights into the efficacy of deploying current compression and quantization techniques on speech foundation models such as Whisper~\cite{whisper}. In addition, statistical significance tests are missing in most existing studies that are important for measuring the efficacy of model compression in terms of ASR performance degradation.

To this end, this paper proposes a novel one-pass multiple ASR systems compression and quantization approach using an all-in-one neural model for both conventional supervised-learning based Conformer and SSL based wav2vec2.0 models. A single system compression cycle allows a large number of weight-sharing target systems (e.g., up to 8) with different compression ratios achieved by using varying Encoder depths, widths, and quantization precision settings to be simultaneously constructed without the need for separate training and storing individual target systems one by one. Experiments conducted using the 300-hr Switchboard dataset trained Conformer and LibriSpeech-100hr fine-tuned wav2vec2.0 SSL models consistently demonstrate the multiple ASR systems being compressed in a single pass produced word error rate (WER) comparable to, or lower by up to {1.01}\% absolute ({6.98}\% relative) than those obtained using individually compressed baseline systems of equal complexity. An overall system compression and training time speed-up ratio of {3.4x} was achieved. Maximum model size compression ratios of {12.8 and 3.93} times were obtained on the two tasks over the 32-bit full-precision baseline Conformer and wav2vec2.0 models, respectively, while incurring non-statistically significant or negligible WER increase.

The main contributions of the paper are summarized below: 
\textbf{1) }To the best of our knowledge, this is the first work that develops an all-in-one neural model for joint architecture compression and parameter quantization of multiple nested Conformer or wav2vec2.0 ASR systems. 
\textbf{2) }The one-pass joint compression-quantization approach can be applied to optimize a rich set of system complexity attributes covering network depth, width, precision, or their combinations. 
\textbf{3) }The efficacy of our one-pass joint compression-quantization approach is demonstrated consistently for both Conformer and wav2vec2.0 models, offering a more comprehensive study than prior research restricted to only one of them~\cite{pru5,jiang2023accurate,lr3,lilossless,rathod2022multi,park2023conformer,quant1,quant2,quant3,qiu2023role,pru4,hj,lodagala2023pada,dist3,quant4}. 
\textbf{4) }Extensive experimental results demonstrate that multiple ASR sub-networks contained in all-in-one models produced WERs comparable to, or lower by up to {1.01}\% absolute ({6.98}\% relative) than, those obtained using individually compressed baseline systems of equal complexity. An overall system compression and training time speed-up ratio of {3.4x} was achieved.

% The rest of this paper are organized as follows: Conformer and wav2vec2.0 ASR models are reviewed in Sections 2 and 3 respectively. Section 4 presents the proposed one-pass multiple system joint compression-quantization framework. Experimental results are reported and analyzed in Section 5. At last, Section 6 concludes and discusses future work. 

\vspace{-0.1cm}
\section{Conformer AED System and Wav2vec2.0 Foundation Model}
\vspace{-0.1cm}
% By integrating convolution neural networks (CNNs) into Transformer architecture~\cite{vaswani2017attention}, a Conformer~\cite{gulati2020conformer} Encoder can capture both local and global dependencies in audio sequences. It comprises multiple stacked blocks, with each block composed of the following modules in sequence with residual connections: a position-wise feed-forward module (FFN), a multi-head self-attention module (MHSA), a convolution module (Conv), and a second FFN module (macaron-like) in the end. In addition, a post-layer normalization is applied to each block. 

% A whole Conformer Encoder layer can be mathematically written as:

% \vspace{-0.3cm}
% \begin{equation}
% \vspace{-0.1cm}
% \addtolength{\jot}{0.2pt}
% \begin{split}
%     &\Tilde{x_i} = x_i + \frac{1}{2}\operatorname{FFN}(x_i); x'_i = \Tilde{x_i} + \operatorname{MHSA}(\Tilde{x_i})\\
%     % &\\
%     &x''_i = x'_i + \operatorname{Conv}(x'_i); y_i = \operatorname{LN}(x'' + \frac{1}{2}\operatorname{FFN}(x''_i))\\
% \end{split},
% \label{eq1}
% \end{equation}
% where $x_i$ and $y_i$ denote the input and output of the $i$-th Encoder layer, respectively. LN$(\cdot)$ is the layer normalization function.

\begin{figure}[th]
    \centering
    \includegraphics[scale=0.4]{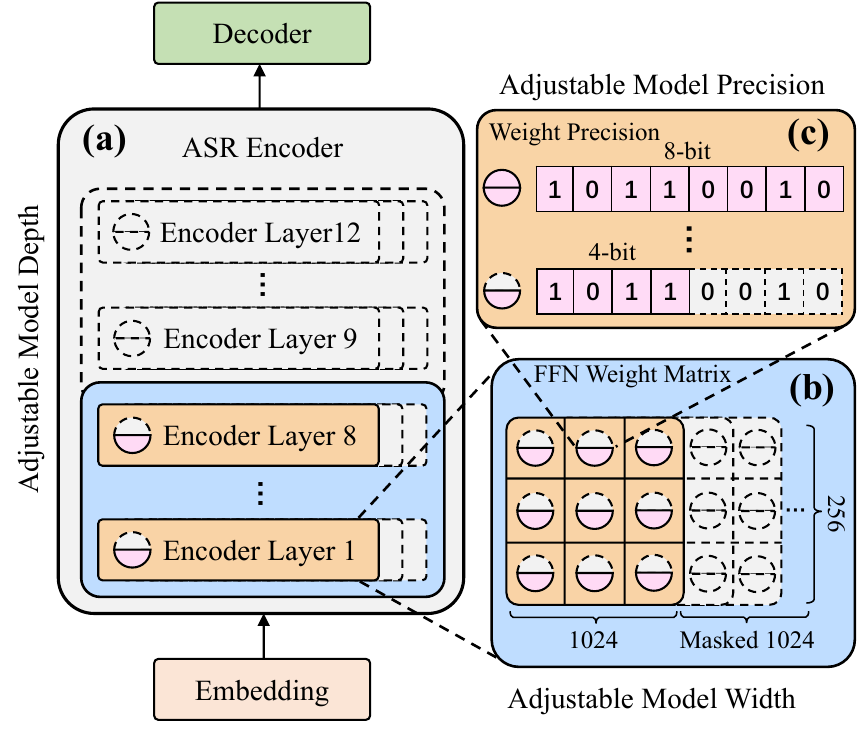}
    \vspace{-0.25cm}
    \caption{Diagram of an all-in-one ASR Encoder. A partly colored area denotes a sub-network, and all sub-networks share weights with larger ones. \textbf{(a)} Sub-networks can skip top layers. \textbf{(b)} Sub-networks can mask partial FFN weight matrices. \textbf{(c)} Sub-networks can be directly quantized to a lower bit-width.}
    \label{fig1}
    \vspace{-0.7cm}
\end{figure}

A Conformer~\cite{gulati2020conformer} Encoder comprises multiple stacked layers. For example, in Fig~\ref{fig1}(a), the Encoder could contain up to 12 layers, which we refer to as \textbf{model depth}. Each Encoder layer consists of the following modules sequentially: a position-wise feed-forward module (FFN), a multi-head self-attention module (MHSA), a convolution module (Conv), and a second FFN module (macaron-like) in the end. Among these modules, FFN accounts for most of the Encoder parameters. Particularly, it consists of two consecutive linear networks, where the first one expands the model dimensionality (e.g., from 256 to 2048, as shown in Fig~\ref{fig1}(b)), and the second one restores it to before. Thus, we refer to \textbf{model width} as the intermediate dimensionality of FFN modules. 

To train an attention-based encoder-decoder (AED) Conformer ASR system, an effective way is using multitask criterion interpolation~\cite{watanabe2017hybrid} between the loss of CTC and the attention Decoder (i.e., a Transformer decoder). This is given by
%%The overall loss function is:
\vspace{-0.2cm}
\begin{equation}
% \scriptsize
    \mathcal{L}_{conformer}=(1-\lambda)\mathcal{L}_{att}+\lambda\mathcal{L}_{ctc},
    \label{eq1}
    \vspace{-0.2cm}
\end{equation}
where $\lambda$ is a constant and empirically set as 0.2
% throughout the experiment s of 
in this paper.

% In contrast to conventional supervised learning methods, the emerging paradigm of self-supervised learning typically trains a foundation model from unlabeled examples to obtain general data representations and fine-tune the model on labeled data. Despite the different training paradigm, speech SSL models such as wav2vec 2.0~\cite{baevski2020wav2vec}, HuBERT~\cite{hsu2021hubert} and WavLM~\cite{chen2022wavlm} share similar Transformer backbones with supervised models. For example, wav2vec2.0 consists of a CNN based feature extractor and a Transformer Encoder, where each Encoder layer is composed of a multi-head self-attention (MHSA) and a position-wise feed-forward (FFN) module.

Speech SSL models such as wav2vec2.0~\cite{baevski2020wav2vec}, HuBERT~\cite{hsu2021hubert}, and WavLM~\cite{chen2022wavlm} share similar Transformer backbones with supervised models. For example, wav2vec2.0 consists of a CNN based feature extractor and a Transformer Encoder, where each Encoder layer contains an MHSA module and an FFN module. In this paper, we fine-tune a pre-trained wav2vec2.0 model with a pure CTC Decoder, namely $\mathcal{L}_{w2v2}=\mathcal{L}_{ctc}$.

In this paper, we perform compression and quantization on the entire Conformer Encoder and the Transformer Encoder excluding the CNNs for wav2vec2.0 models.

\vspace{-0.1cm}
\section{One-Pass Multiple Systems Compression}
\vspace{-0.1cm}
% In this section, we presents the proposed all-in-one neural model for both Conformer and SSL Transformer Encoder, which enables adjustable architecture hyper-parameters including network depth, width and weight precision. We use one-pass multitask training method to obtain multiple target systems within one compressing cycle. The Kullback–Leibler (KL) divergence regularization is adopted to further enhance the model performance.

\subsection{All-In-One Weight-Sharing Neural Model}
\vspace{-0.1cm}
% In this work, an all-in-one neural network contains various sub-networks with different depths, widths and quantization precision settings, as shown in Fig.~\ref{fig1}. Once an all-in-one neural model is trained, it takes no retraining and storage overhead to compress the model to a smaller one as they share the same weights. For example, one can obtain a 4-bit 8-layer compressed model from a 8-bit 12-layer all-in-one model by directly truncating the top 4 layers and extra weight resolutions without needs of retraining. Now we describe in detail the design of each architecture dimension.

Similar to once-for-all techniques~\cite{chen2023once,yu2020bignas,hu2022generalizing}, an all-in-one model refers to a neural network containing various sub-networks, where all small networks share parameters with larger ones. For example, in Fig.~\ref{fig1}, an 8-layer 1024-dim 4-bit small model is nested in the 12-layer 2048-dim 8-bit all-in-one model. Once an all-in-one model is trained, the target sub-networks are automatically contained, resulting in no retraining and storage overhead to compress the model to a smaller one. In this paper, we propose an all-in-one model that offers flexibility in model depths, widths, quantization precision, and their combinations for both Conformer and wav2vec2.0.

\vspace{-0.25cm}
\subsection{Multiple Weight-Sharing Systems Training}
\vspace{-0.1cm}
To obtain an all-in-one neural network, a one-pass multitask training approach is needed to ensure all the sub-networks are also well trained. The training criterion is given by
\vspace{-0.25cm}
\begin{equation}
    \mathcal{L}=\mathcal{L}_{max} + \sum_{i\in\mathcal{M}}\lambda^1_{i}\mathcal{L}_i
    \label{eq2}
    \vspace{-0.3cm}
\end{equation}
where $\mathcal{L}_{max}$ is the loss of the largest network, $\mathcal{M}$ is the set of other target sub-network structures, and $\mathcal{L}_{i}$ is the loss of sub-network $i$, with a constant coefficient $\lambda^1_{i}$. In Eq.~\ref{eq2}, the first term is exactly the same as conventional single system training, and the second term is to enable sub-networks. Note that in our default setup, the largest model is the same size as our chosen baselines since a major goal of this paper is model compression.

Since all sub-networks are nested in larger ones, their losses are obtained by forwarding through only a partial of the largest network. Specifically, when compressing through depth, the several top layers can be skipped (Fig.~\ref{fig1}(a)). For width, a certain portion of rows (or columns) of FFN weight matrices can be masked\footnote{For wav2vec2.0, rows (columns) having small $L_2$-norm values are considered of less importance and masked with high-priority.} with zero before forwarding (Fig.~\ref{fig1}(b)). For quantization, the weights can be directly quantized to the target bit-width before forwarding (Fig.~\ref{fig1}(c)). Therefore, with such methods, we only need to store the largest network in the end while multiple sub-networks are already automatically contained.

\vspace{-0.25cm}
\subsection{Kullback–Leibler Divergence Regularization}
\vspace{-0.1cm}
Like the effect of knowledge distillation, when simultaneously training multiple systems, larger models are implicitly providing guidance to smaller ones, while we can make this effect explicit by adding an extra Kullback–Leibler (KL) divergence regularization. Denote $p_{max}$, and $p_{i}, (i\in\mathcal{M}$) as the output distribution of the largest network and sub-network $i$, respectively. The KL regularization term is given by
\vspace{-0.3cm}
\begin{equation}
    \Omega_{i} = D_{KL}(SG(p_{max})||p_{i}),
    \label{eq3}
    \vspace{-0.3cm}
\end{equation}
where $SG(\cdot)$ denotes the stop-gradient operation preventing the gradient from flowing back to the largest structure. Finally, we can write the criterion with KL regularization:
\vspace{-0.3cm}
\begin{equation}
    \mathcal{L}_{kl} = \mathcal{L}_{max} + \sum_{i\in\mathcal{M}}(\lambda^1_i\mathcal{L}_{i} + \lambda^2_i\Omega_{i}),
    \label{eq4}
    \vspace{-0.4cm}
\end{equation}
where $\lambda^1_i$ and $\lambda^2_i$ are constant coefficients.

\vspace{-0.3cm}
\section{Experiments}
\vspace{-0.1cm}
\subsection{Experimental Setup}
\vspace{-0.1cm}
\noindent\textbf{Baseline Models.}
The baseline Conformer system is configured using the ESPnet~\cite{watanabe2018espnet} recipe\footnote{\href{https://github.com/espnet/espnet/tree/master/egs/swbd/asr1}{ESPnet: egs/swbd/asr1/run.sh}}. For wav2vec2.0 models, the wav2vec2-base-100h is downloaded from Huggingface\footnote{\href{https://huggingface.co/facebook/wav2vec2-base-100h}{Huggingface: facebook/wav2vec2-base-100h}}. We fine-tuned wav2vec2-base-100h for 5 epochs as our baseline. %we compress \red{wav2vec2-base} pre-trained model on HuggingFace Transformer.

\noindent\textbf{Data.}
The Conformer systems are trained with 300-hr benchmark Switchboard corpus~\cite{godfrey1992switchboard} and evaluated on NIST Hub5’00, RT02, and RT03 evaluation sets. We fine-tune the wav2vec2-base-100h model on LibriSpeech’s 100-hour clean subset~\cite{panayotov2015librispeech}.

\noindent\textbf{Training.}
All the Conformer systems, including one-pass and individually compressed ones, are trained from scratch with exactly the same learning schedule as the baseline Conformer recipe. For wav2vec2.0 systems, we fine-tune wav2vec2-base-100h with a batch size of around 202 seconds of audios for 5 epochs, which is consistent with our baseline. We use the AdamW optimizer with a learning rate of 3e-5. A linear warmup is used for the first 10\% training steps, followed by a linear decay to zero. All the experiments in this paper are conducted on a single NVIDIA A40 (48G) GPU. 

\noindent\textbf{Quantization.} In this paper, we apply symmetric quantization on model weights using quantization-aware training (QAT) with a straight-through estimator and a learnable scale factor~\cite{nagel2021white}.

\vspace{-0.25cm}
\subsection{Results of Conformer Systems}
\vspace{-0.1cm}
\textbf{Single Attribute Compression:} We first conduct experiments on Conformer systems with all-in-one models sharing weights through each network attribute separately. We study two configurations for each attribute, i.e., \{8, 12\} layers for model depth, \{1024, 2048\} for model width, and \{4-bit, 8-bit\} for weight precision. Results evaluated on Hub5’00, RT02, and RT03 sets are shown in Tab.~\ref{tab1}. For a clearer comparison, we also reported the average WER over all three test sets, where several trends can be found: \textbf{1)} On whichever attribute, all the sub-networks contained in all-in-one models produce significantly lower word error rate (WER) than individually compressed systems of equal network architecture. (e.g., in Tab.~\ref{tab1}, Sys.~2$s$ and 3$s$ vs. Sys.~1; Sys.~9$s$ vs. Sys.~7, etc.). \textbf{2)} On whichever attribute, the largest sub-network (with the same size as the baseline) of all-in-one model produces significantly lower WER than the baseline system (i.e., Sys.~2$\ell$,3$\ell$,5$\ell$,6$\ell$,9$\ell$,10$\ell$ vs. Sys.~0 in Tab.~\ref{tab1}). Remarkably, some small sub-networks even have performance better than (i.e., Sys.~6$s$ and 10$s$ vs. Sys.~0, in Tab.~\ref{tab1}) or comparable to (i.e., Sys.~5$s$ and 9$s$ vs. Sys.~0, in Tab.~\ref{tab1}) the baseline. \textbf{3)} On whichever attribute, the KL regularization\footnote{For Conformer systems, we set $\lambda^1=0.8$ and $\lambda^2=0.2$ for depth, and $\lambda^1=\lambda^2=1$ for other attributes.} further improves the performance of all-in-one models (e.g., Sys.~3$\ell$ vs. Sys.~2$\ell$; Sys.~6$s$ vs. Sys.~5$s$, etc.). 

\noindent\textbf{Multiple Attribute Compression:}
On top of compressing separately, multiple attribute compression of Conformer systems with all-in-one models is studied in Tab.~\ref{tab2}, where the architecture configuration list is the combination of attributes in Tab.~\ref{tab1}, and different attributes can be compressed simultaneously. For Conformer all-in-one models, when training, we only forward through 3 sub-networks (i.e., the largest and smallest one, and another one sampled from the other sub-networks) in one updating iteration instead of all of them to speed up the training process. Also, we average all the test sets for clearer comparisons. Several trends can be found: \textbf{1)} All the sub-networks in the all-in-one model consistently produce significantly lower WER than that of individually compressed systems of equal complexity (i.e., Sys.~1A$\sim$8A vs. Sys.~1$\sim$8 in Tab.~\ref{tab2}, respectively). \textbf{2)} Also, further performance improvements (or WER decrease) can be found when adding KL regularization, especially for small sub-networks (i.e., Sys.~1K$\sim$8K vs. Sys.~1A$\sim$8A in Tab.~\ref{tab2}, respectively). \textbf{3)} With one-pass multitask training and KL regularization, a WER reduction up to 1.01\% absolute (6.98\% relative) is achieved for the smallest sub-network of the architecture 8-1024-4bit against the individually compressed system of the same architecture (i.e., Avg. of Sys.~8K vs. Sys.~8 in Tab.~\ref{tab2}). Moreover, a maximum compression ratio of 12.8 times is obtained against the baseline Conformer system while incurring no statistically significant WER increase (i.e., Avg. of Sys.~4K vs. Sys.~0 in Tab.~\ref{tab2}). \textbf{4)} An overall compression and training time speed-up ratio of 3.4 times is achieved (see Training time in Tab.~\ref{tab2}, where the overall training time is measured by adding up the GPU hours needed for training all the sub-networks, namely 319 hours in total for training individual sub-networks separately). The all-in-one Conformer Encoder only contains 34 million parameters, while it can be re-configured to 8 different-sized systems that would contain 186m parameters in total if trained and stored separately (see \#Param. in Tab.~\ref{tab2}).

\begin{table}[]
\caption{WER($\downarrow$) of all-in-one Conformer models against individually compressed Conformer systems through a single network attribute. Systems with the same Sys. Number share weights, where ``$\ell$" and ``$s$" are for the largest and small sub-network, respectively. Blue cells: the performance is better than the baseline (Sys.~0) with a statistical significance (MAPSSWE~\cite{gillick1989some}, $\alpha$=0.05). Yellow cells: the performance has no significant difference from the baseline. * means it is significantly better than the individually compressed system of equal model size (Sys.~1,4,7, and 8).}
\vspace{-0.25cm}
    \centering
    \setlength\tabcolsep{1pt}
    \resizebox{\linewidth}{!}{
    \begin{tabular}{l|c|c|ccc|ccccccc|c}
        \hline
        \hline
        \multirow{2}{*}{Sys.} & \multicolumn{2}{c|}{\multirow{2}{*}{Attribute}} & \multirow{2}{*}{\shortstack{All-\\in-one}} & \multicolumn{1}{|c}{\multirow{2}{*}{\shortstack{+\\KL}}} & \multicolumn{1}{|c|}{\multirow{2}{*}{\shortstack{\#\\Parm.}}}&\multicolumn{2}{c}{Hub5’00} & \multicolumn{3}{|c|}{RT02} & \multicolumn{2}{c|}{RT03} & {\multirow{2}{*}{Avg.}}\\
        \cline{7-13}
         & \multicolumn{2}{c|}{} & &\multicolumn{1}{|c}{} & \multicolumn{1}{|c|}{}& swbd & calhm & \multicolumn{1}{|c}{swbd1} & swbd2 & \multicolumn{1}{c|}{swbd3} & swbd & fsh & \\
         \hline
         \hline
         0 & \multicolumn{4}{c|}{12-2048-32bit (baseline)} &34m & 7.4 & 15.2 & 8.9 & 13.0 & 15.9 & 10.6 & 16.6 & 12.86\\
         \hline
         \hline
         1 & \multirow{5}{*}{Dep.} & 8 & \xmark & - &23m& 7.8  & 15.9  & 9.7 & 13.4 & 16.5 & 11.2 & 17.5 & 13.49\\
         \cline{3-5}\cline{6-6}
         2$s$ & & 8 & \multirow{2}{*}{\cmark} & \multirow{2}{*}{\xmark} &share& 7.8 & 15.6 & 9.2 & 13.7 & 16.4 & 10.7 & 17.0 & 13.25$^*$\\
         2$\ell$ & & 12 & & &34m& 7.4 & 15.0 & 9.0 & 13.1 & 15.8 & 10.3 & 16.3 & {\cellcolor{cyan!15}12.71$^*$}\\
         \cline{3-3}\cline{4-6}
         3$s$ & & 8 & \multirow{2}{*}{\cmark} & \multirow{2}{*}{\cmark} &share& 7.8 & 15.8 & 9.4 & 13.3 & 16.5 & 10.8 & 16.9 & 13.23$^*$\\
         3$\ell$ & & 12 & & &34m& 7.3 & 14.9 & 8.8 & 12.7 & 15.7 & 10.4 & 16.2 & {\cellcolor{cyan!15}12.59$^*$}\\
         \hline
         \hline
         4 & \multirow{5}{*}{Wid.} & 1024 & \xmark & - &21m& 7.7  & 15.5 & 9.0 & 13.3 & 16.2 &10.8 & 16.9 & 13.06\\
         \cline{3-5}\cline{6-6}
         5$s$ & & 1024 & \multirow{2}{*}{\cmark} & \multirow{2}{*}{\xmark} &share&7.3&15.1&9.0&13.1&15.8&10.6&16.9&{\cellcolor{orange!20}12.88$^*$} \\
         5$\ell$ & & 2048 & & &34m& 7.1 & 14.8 &8.7&12.6&15.5&10.3&16.5&{\cellcolor{cyan!15}12.54$^*$} \\
         \cline{3-3}\cline{4-6}
         6$s$ & & 1024 & \multirow{2}{*}{\cmark} & \multirow{2}{*}{\cmark} &share& 7.4 &15.0&8.8&12.7&15.2&10.2&16.4&{\cellcolor{cyan!15}12.54$^*$} \\
         6$\ell$ & & 2048 & & &34m& 7.3 &14.8&8.5&12.5&14.9&10.0&16.1&{\cellcolor{cyan!15}12.33$^*$} \\
         \hline
         \hline
         7 & \multirow{6}{*}{Prec.} & 4-bit & \multirow{2}{*}{\xmark} & \multirow{2}{*}{-} &34m& 7.6 & 15.6 & 8.9&13.3&16.1&10.6&17.0&13.06  \\
         8 & & 8-bit & & &34m& 7.4 &15.2 &8.9&13.0&15.8&10.6&16.7&12.86 \\
         \cline{3-5}\cline{6-6}
         9$s$ & & 4-bit & \multirow{2}{*}{\cmark} & \multirow{2}{*}{\xmark} &share& 7.5 &15.5 &9.0&12.9&15.6&10.5&16.9&{\cellcolor{orange!20}12.90$^*$} \\
         9$\ell$ & & 8-bit & & &34m& 7.4 &15.2 &8.8&12.6&15.5&10.4&16.6&{\cellcolor{cyan!15}12.70$^*$} \\
         \cline{3-3}\cline{4-6}
         10$s$ & & 4-bit & \multirow{2}{*}{\cmark} & \multirow{2}{*}{\cmark} &share& 7.0 &15.2 &8.8&12.6&15.5&10.1&16.3&{\cellcolor{cyan!15}12.54$^*$} \\
         10$\ell$ & & 8-bit & & &34m& 7.1 & 15.0& 8.8&12.6&15.6&10.1&16.3&{\cellcolor{cyan!15}12.53$^*$}\\
         \hline
         \hline

    \end{tabular}
        }
    \label{tab1}
    \vspace{-0.7cm}

\end{table}

\begin{table*}[]
\caption{WER($\downarrow$) of all-in-one Conformer models against individually compressed Conformer systems through multiple network attributes. Systems with the same letter (A or K) share weights; systems with the same number (1$\sim$8) share the same network architecture. Blue cells: the performance is better than the baseline (Sys.~0) with a statistical significance (MAPSSWE~\cite{gillick1989some}, $\alpha$=0.05). Yellow cells: the performance has no significant difference from the baseline. * means it is significantly better than the individually compressed system of equal model size (Sys.~1$\sim$8). $\dag$ means it has no significant difference with the corresponding individually-compressed system.}
    \centering
    \vspace{-0.25cm}
    \setlength\tabcolsep{10pt}
    \resizebox{\linewidth}{!}{
    \begin{tabular}{l|c|c|c|c|c|c|ccccccc|c}
        \hline
        \hline
        \multirow{2}{*}{Sys.} & \multirow{2}{*}{Architecture} & \multirow{2}{*}{\shortstack{All-\\in-one}} & \multirow{2}{*}{+KL} & \multirow{2}{*}{\shortstack{\#Parameters\\Encoder}}& \multirow{2}{*}{\shortstack{Training \\ Time}} & \multirow{2}{*}{\shortstack{Compression\\Ratio}} & \multicolumn{2}{c}{Hub5’00} & \multicolumn{3}{|c|}{RT02} & \multicolumn{2}{c|}{RT03} & {\multirow{2}{*}{Avg.}}\\
        \cline{8-14}
         & & & & & & &swbd & calhm & \multicolumn{1}{|c}{swbd1} & swbd2 & \multicolumn{1}{c|}{swbd3} & swbd & fsh & \\
         \hline
         \hline
         0 & \multicolumn{3}{c|}{12-2048-32bit (baseline)} &34m &37hrs & 1x & 7.4 & 15.2 & 8.9 & 13.0 & 15.9 & 10.6 & 16.6 & 12.86\\
         \hline
         \hline
         1 & 12-2048-8bit & \multirow{8}{*}{\xmark} & \multirow{8}{*}{-} &34m &48hrs & 8.0x & 7.4 &15.2 &8.9&13.0&15.8&10.6&16.7&12.86\\
         2 & 12-2048-4bit & & &34m &48hrs & 4.0x & 7.6 & 15.6 & 8.9&13.3&16.1&10.6&17.0&13.06  \\
         3 & 12-1024-8bit & & &21m &42hrs & 6.4x & 7.9 &15.7 &9.1&13.6&16.4 &10.7&17.5&13.33  \\
         4 & 12-1024-4bit & & &21m &42hrs & 12.8x & 7.8 &16.0 &9.5&13.4&16.7&10.9&17.5&13.46  \\
         5 & 8-2048-8bit & & &23m &37hrs & 5.8x & 8.0 &15.7 &9.3&13.5&16.6&11.0&17.4 &13.49  \\
         6 & 8-2048-4bit & & &23m &37hrs & 11.7x & 8.1 &16.2 &9.6&14.1&17.3&11.4&18.3 &13.96  \\
         7 & 8-1024-8bit & & &15m &33hrs & 8.2x & 7.9 &16.1 &9.5&14.9&16.8&11.2&17.8 &13.69  \\
         8 & 8-1024-4bit & & &15m &32hrs & 18.4x & 8.4 &17.3 &10.2&14.7&17.7&11.9&18.7 &14.48  \\
         \hline
         \hline
         1A & 12-2048-8bit & \multirow{8}{*}{\cmark} & \multirow{8}{*}{\xmark} & \multirow{8}{*}{\shortstack{share\\34m}} &\multirow{8}{*}{94hrs} & 4.0x & 7.2 &15.1 &8.7&12.7&15.3&10.2&16.5 &{\cellcolor{cyan!15}12.58$^*$} \\
         2A & 12-2048-4bit & & & & & 8.0x & 7.3 &15.2 &9.0&13.0&15.5&19.4&16.8 &{\cellcolor{orange!20}12.80$^*$}  \\
         3A & 12-1024-8bit & & & & & 6.4x & 7.7 &15.3 &9.2&13.0&16.0&10.5&17.3 &13.06$^*$  \\
         4A & 12-1024-4bit & & & & & 12.8x & 7.7 &15.6 &9.3&13.4&16.3&10.6&17.3 &13.23$^*$  \\
         5A & 8-2048-8bit & & & & & 5.8x & 7.8 &15.6 &9.4&13.4&16.2&10.7&17.4 &13.28$^*$  \\
         6A & 8-2048-4bit & & & & & 11.7x & 7.9 &16.1 &9.7&13.9&16.6&10.9&17.5 &13.54$^*$  \\
         7A & 8-1024-8bit & & & & & 8.2x & 8.0 &16.1 &9.6&13.9&16.8&11.1&17.9 &13.68$^\dag$  \\
         8A & 8-1024-4bit & & & & & 18.4x & 8.2 &16.1 &9.8&14.2&17.2&11.2&18.1 &13.90$^*$  \\
         \hline
         \hline
         1K & 12-2048-8bit & \multirow{8}{*}{\cmark} & \multirow{8}{*}{\cmark} & \multirow{8}{*}{\shortstack{share\\34m}} &\multirow{8}{*}{95hrs} & 4.0x & 7.2 &14.8 &8.9&12.6&15.4&10.2&16.3 &{\cellcolor{cyan!15}12.52$^*$} \\
         2K & 12-2048-4bit & & & & & 8.0x & 7.2 &15.0 &8.9&12.9&15.5&10.3&16.5 & {\cellcolor{cyan!15}12.66$^*$}  \\
         3K & 12-1024-8bit & & & & & 6.4x & 7.5 &15.2 &9.1&12.8&15.9&10.3&16.4 &{\cellcolor{orange!20}12.76$^*$}  \\
         4K & 12-1024-4bit & & & & & 12.8x & 7.6 &15.4 &9.1&13.0&16.0&10.6&16.8 &{\cellcolor{orange!20}12.97$^*$}  \\
         5K & 8-2048-8bit & & & & & 5.8x & 7.6 &15.3 &9.2&13.3&16.3&10.7&16.8 &13.04$^*$  \\
         6K & 8-2048-4bit & & & & & 11.7x & 7.7 &15.4 &9.2&13.5&16.4&10.8&17.0 &13.19$^*$  \\
         7K & 8-1024-8bit & & & & & 8.2x & 7.8 &15.8 &9.4&13.4&16.5&10.8&17.3 &13.32$^*$  \\
         8K & 8-1024-4bit & & & & & 18.4x & 7.9 &15.8 &9.5&13.6&16.7&11.0&17.5 &13.47$^*$  \\
         \hline
         \hline

    \end{tabular}
        }
    
    \label{tab2}
    \vspace{-0.4cm}

\end{table*}

\begin{table}[ht]
\caption{WER($\downarrow$) of all-in-one wav2vec2.0 models against individually compressed wav2vec2.0 systems through a single network attribute. Systems with the same Sys. Number share weights, where ``$\ell$" and ``$s$" are for the largest and small sub-network, respectively.  Colors and marks are with the same meanings as the MAPSSWE~\cite{gillick1989some} test in Table~\ref{tab1}.}
    \centering
    \vspace{-0.3cm}
    \setlength\tabcolsep{7pt}
    \resizebox{\linewidth}{!}{
    \begin{tabular}{l|c|c|ccc|cccc}
        \hline
        \hline
        \multirow{2}{*}{Sys.} & \multicolumn{2}{c|}{\multirow{2}{*}{Attribute}} & \multirow{2}{*}{\shortstack{All-\\in-one}} & \multicolumn{1}{|c}{\multirow{2}{*}{\shortstack{+\\KL}}} &\multicolumn{1}{|c|}{\multirow{2}{*}{\shortstack{\#\\Parm.}}}& \multicolumn{2}{c}{dev} & \multicolumn{2}{c}{test}\\
        \cline{7-10}
         & \multicolumn{2}{c|}{} &&\multicolumn{1}{|c}{}& \multicolumn{1}{|c|}{}&  clean & other & clean & other\\
         \hline
         \hline
         0 & \multicolumn{4}{c}{12-3072-32bit (baseline)} & \multicolumn{1}{|c|}{94m}&5.78 & 13.65 & 5.89 & 13.32\\
         \hline
         \hline
         1 & \multirow{5}{*}{Depth} & 11 & \xmark & - & 87m & {5.88}   & 13.96   & {5.94}   & 13.54 \\
         \cline{3-6}
         2$s$ & & 11 & \multirow{2}{*}{\cmark} & \multirow{2}{*}{\xmark} & share & {\cellcolor{orange!20}5.88$^{\dag}$}  &	14.10$^{\dag}$ &	{\cellcolor{orange!20}5.98$^{\dag}$} &	13.62$^{\dag}$ \\
         2$\ell$ & & 12 & & & 94m & {\cellcolor{orange!20}5.79$^\dag$}   & 13.87   & {\cellcolor{orange!20}5.82$^\dag$}   & {\cellcolor{orange!20}13.31$^\dag$}\\
         \cline{3-6}
         3$s$ & & 11 & \multirow{2}{*}{\cmark}  & \multirow{2}{*}{\cmark} &share &5.95$^{\dag}$ &	14.00$^{\dag}$ &	{\cellcolor{orange!20}5.89$^{\dag}$} &	13.59$^{\dag}$\\
         3$\ell$ & & 12 & & & 94m & {\cellcolor{orange!20}5.73$^\dag$} 	&{\cellcolor{orange!20}13.81$^\dag$} &	{\cellcolor{cyan!15}5.76$^*$} 	&{\cellcolor{orange!20}13.31$^\dag$}\\
         \hline
         \hline
         4 & \multirow{5}{*}{Width} & 2560 & \xmark & - &85m&6.58 	&15.41 &	6.55 &	14.99\\
         \cline{3-6}
         5$s$ & & 2560 & \multirow{2}{*}{\cmark} & \multirow{2}{*}{\xmark} & share & 6.47$^{\dag}$ &	15.62 &	6.56$^{\dag}$ &	15.27\\
         5$\ell$ & & 3072 & & & 94m &{\cellcolor{orange!20}5.81$^\dag$} &	{\cellcolor{cyan!15}13.32$^*$} &	{\cellcolor{cyan!15}5.77$^*$} &	{\cellcolor{cyan!15}13.00$^*$}\\
         \cline{3-6}
         6$s$ & & 2560 &  \multirow{2}{*}{\cmark} & \multirow{2}{*}{\cmark} & share & 6.37$^*$ &	15.23$^{\dag}$ &	6.32$^*$ 	&14.82$^{\dag}$\\
         6$\ell$ & & 3072 & & & 94m & {\cellcolor{orange!20}5.78$^\dag$} &	{\cellcolor{orange!20}13.60$^\dag$} &	{\cellcolor{cyan!15}5.75$^*$} &	{\cellcolor{orange!20}13.12$^\dag$}\\
         \hline
         \hline
         7 & \multirow{6}{*}{Prec.} & 6-bit & \multirow{2}{*}{\xmark} & \multirow{2}{*}{-} & 94m & {5.86} &	{13.85} &	{5.85} &	{13.30}  \\
         8 & & 8-bit & & & 94m &{5.80} 	&{13.63} 	&{5.83} &	{13.26}\\
         \cline{3-6}
         9$s$ & & 6-bit & \multirow{2}{*}{\cmark} & \multirow{2}{*}{\xmark}  &share &{\cellcolor{orange!20}5.84$^{\dag}$} 	&13.96$^{\dag}$ 	&{\cellcolor{orange!20}5.88$^{\dag}$} 	&{\cellcolor{orange!20}13.40$^{\dag}$}\\
         9$\ell$ & & 8-bit & & &94m & {\cellcolor{orange!20}5.76$^{\dag}$} &	{\cellcolor{orange!20}13.62$^{\dag}$} &	{\cellcolor{cyan!15}5.75$^\dag$} &	{\cellcolor{cyan!15}13.08$^*$}\\
         \cline{3-6}
         10$s$ & & 6-bit & \multirow{2}{*}{\cmark}  & \multirow{2}{*}{\cmark} &share &{\cellcolor{orange!20}5.72$^*$} 	&{\cellcolor{orange!20}13.46$^*$}	&{\cellcolor{cyan!15}{5.75}$^{\dag}$} 	&{\cellcolor{orange!20}13.21$^{\dag}$}\\
         10$\ell$ & & 8-bit & & & 94m &{\cellcolor{orange!20}5.81$^{\dag}$} 	&{\cellcolor{orange!20}13.63$^{\dag}$} 	&{\cellcolor{cyan!15}5.71$^{\dag}$} 	&{\cellcolor{orange!20}13.19$^{\dag}$}\\
         \hline
         \hline

    \end{tabular}
        }
    
    \label{tab3}
    \vspace{-0.7cm}

\end{table}

\begin{table}[ht]
\caption{WER($\downarrow$) of all-in-one wav2vec2.0 models against individually compressed wav2vec2.0 systems through multiple network attributes. Systems with the same letter (A or K) share weights, while systems with the same Sys. Number (1$\sim$8) share the same network architecture. Colors and marks are with the same meanings as the MAPSSWE~\cite{gillick1989some} test in Table~\ref{tab2}.}
\vspace{-0.3cm}
    \centering
    \setlength\tabcolsep{3pt}
    \resizebox{\linewidth}{!}{
    \begin{tabular}{l|c|c|c|c|c|c|cccc}
        \hline
        \hline
        \multirow{2}{*}{Sys.} & \multirow{2}{*}{Architecture} & \multirow{2}{*}{\shortstack{All-\\in-one}} & \multirow{2}{*}{+KL} & \multirow{2}{*}{\shortstack{\#\\Parm.}} & \multirow{2}{*}{\shortstack{Train\\Time}} & \multirow{2}{*}{\shortstack{Comp.\\Ratio}} & \multicolumn{2}{c}{dev} & \multicolumn{2}{c}{test}\\
        \cline{8-11}
         & & & & & & &  clean & other & clean & other\\
         \hline
         \hline
         0 & \multicolumn{3}{c|}{12-3072-32bit (baseline)} & 94m & 3.8hrs & 1x & 5.78 & 13.65 & 5.89 & 13.32\\
         \hline
         \hline
         1 & 12-3072-8bit & \multirow{8}{*}{\xmark} & \multirow{8}{*}{-} & 94m & 4.0hrs & 3.08x &{5.80} 	&{13.63} 	&{5.83} &	{13.26} \\
         2 & 12-3072-6bit & & & 94m & 4.1hrs & 3.72x & {5.86} &	{13.85} &	{5.85} &	{13.30}\\
         3 & 12-2560-8bit & & & 85m & 4.2hrs & 3.34x &6.53 &	15.19 &	6.59 &	14.96\\
         4 & 12-2560-6bit & & & 85m & 4.0hrs & 4.00x & 6.62 &	15.68 &	6.67 &	15.24\\
         5 & 11-3072-8bit & & & 87m & 4.0hrs & 3.27x &5.93 &	{13.85} &	{5.99} &	{13.33}\\
         6 & 11-3072-6bit & & & 87m & 4.0hrs & 3.93x & 6.01 &	14.04 &	6.03 &	13.68\\
         7 & 11-2560-8bit & & & 79m & 3.9hrs & 3.53x &6.68 	&15.41 &	6.70 &	15.14\\
         8 & 11-2560-6bit & & & 79m & 4.0hrs & 4.22x &6.83 	&16.00 &	6.85 &	15.63\\
         \hline
         \hline
         1A & 12-3072-8bit & \multirow{8}{*}{\cmark} & \multirow{8}{*}{\xmark} & \multirow{8}{*}{\shortstack{share\\94m}} & \multirow{8}{*}{22.9hrs} & 3.08x &{\cellcolor{cyan!15}5.62$^*$} &	{\cellcolor{cyan!15}13.43$^{\dag}$} &	{\cellcolor{cyan!15}5.61$^*$} 	&{\cellcolor{cyan!15}12.94$^*$}\\
         2A & 12-3072-6bit & & & & & 3.72x &{\cellcolor{orange!20}5.65$^*$} &	{\cellcolor{orange!20}13.60$^*$} &	{\cellcolor{cyan!15}5.69$^*$} &	{\cellcolor{orange!20}13.18$^{\dag}$}\\
         3A & 12-2560-8bit & & & & & 3.34x &6.08$^*$ 	&14.98$^*$ &	6.13$^*$ &	14.48$^*$\\
         4A & 12-2560-6bit & & & & & 4.00x &6.08$^*$ &	15.21$^*$ &	6.17$^*$ &	14.70$^*$\\
         5A & 11-3072-8bit & & & & & 3.27x &{\cellcolor{orange!20}5.71$^*$} &	{\cellcolor{orange!20}13.66$^{\dag}$} &	{\cellcolor{orange!20}5.78$^*$} &	{\cellcolor{orange!20}13.20$^{\dag}$}\\
         6A & 11-3072-6bit & & & & & 3.93x  &{\cellcolor{orange!20}5.77$^*$}&	{\cellcolor{orange!20}13.86$^{\dag}$}& 	{\cellcolor{orange!20}5.86$^*$}&	{\cellcolor{orange!20}13.46$^*$}\\
         7A & 11-2560-8bit & & & & & 3.53x &6.18$^*$ &	15.21$^{\dag}$ &	6.24$^*$ &	14.68$^*$\\
         8A & 11-2560-6bit & & & & & 4.22x &6.23$^*$ &	15.43$^*$ &	6.35$^*$ &	14.99$^*$\\
         \hline
         \hline
         1K & 12-3072-8bit & \multirow{8}{*}{\cmark} & \multirow{8}{*}{\cmark} & \multirow{8}{*}{\shortstack{share\\94m}} & \multirow{8}{*}{23.1hrs} & 3.08x &{\cellcolor{cyan!15}5.62$^*$} &	{\cellcolor{cyan!15}13.33$^*$} &	{\cellcolor{cyan!15}5.60$^*$} &	{\cellcolor{cyan!15}13.12$^{\dag}$}\\
         2K & 12-3072-6bit & & & & & 3.72x & {\cellcolor{orange!20}5.65$^*$}& 	{\cellcolor{orange!20}13.50$^*$} &	{\cellcolor{cyan!15}5.67$^*$} &	{\cellcolor{orange!20}13.16$^{\dag}$}\\
         3K & 12-2560-8bit & & & & & 3.34x &6.03$^*$ 	&14.86$^*$& 	6.07$^*$ &	14.44$^*$\\
         4K & 12-2560-6bit & & & & & 4.00x &6.08$^*$ &	14.95$^*$ &	6.09$^*$ &	14.54$^*$\\
         5K & 11-3072-8bit & & & & & 3.27x &{\cellcolor{orange!20}5.74$^*$} &	{\cellcolor{orange!20}13.57$^*$}& 	{\cellcolor{orange!20}5.78$^*$} &	{\cellcolor{orange!20}13.35$^{\dag}$}\\
         6K & 11-3072-6bit & & & & & 3.93x &{\cellcolor{orange!20}5.77$^*$}	&{\cellcolor{orange!20}13.71$^*$} &	{\cellcolor{orange!20}5.81$^*$}&	{\cellcolor{orange!20}13.42$^*$}\\
         7K & 11-2560-8bit & & & & & 3.53x &6.14$^*$ &	15.06$^*$ &	6.22$^*$ &	14.73$^*$\\
         8K & 11-2560-6bit & & & & & 4.22x & 6.23$^*$ &	15.15$^*$ &	6.26$^*$ &	14.75$^*$\\
         \hline
         \hline

    \end{tabular}
        }
    
    \label{tab4}
    \vspace{-0.5cm}

\end{table}

\vspace{-0.3cm}
\subsection{Results for wav2vec2.0 Systems}
\vspace{-0.1cm}
We then demonstrate the efficacy of the one-pass all-in-one model on SSL systems. For a fair comparison, we fine-tune individual and all-in-one models with the same iterations.

\noindent\textbf{Single Attribute Compression:}
We first study single attribute compression for wav2vec2.0 systems, where the configuration of each attribute is set as \{11, 12\} for model depth, \{2560, 3072\} for model width, and \{6-bit, 8-bit\} for weight precision. Results are shown in Tab.~\ref{tab3}. Similar trends can be found to those observed in Conformer systems. \textbf{1)} Sub-networks in all-in-one models have performance of no significant differences with (e.g., Sys.~3$s$ and 2$s$ vs. Sys.~1, etc.) or even better than (e.g., Sys.~5$\ell$ vs. Sys.~0; Sys.~10$s$ vs. Sys.~7, etc.) individual systems. \textbf{2)} They are further enhanced by KL regularization\footnote{For SSL cases, we empirically set $\lambda^1=1$ and $\lambda^2=0.005$ for depth; $\lambda^1=1$ and $\lambda^2=0.1$ for other attributes.} (e.g., Sys.~6$s$ vs. Sys.~5$s$, Sys.~10$s$ vs. Sys.~9$s$, etc.).

\noindent\textbf{Multiple Attribute Compression:} Experiments of multiple attribute compression of wav2vec2.0 systems also share trends with those in Conformer systems found in Tab.~\ref{tab2}. The results are reported in Tab.~\ref{tab4}. Particularly, \textbf{1)} a WER reduction up to 0.88\% absolute (test other of Sys.~8K vs. Sys.~8 in Tab.~\ref{tab4}) and 8.8\% relative (dev clean of Sys.~8K vs. Sys.~8 in Tab.~\ref{tab4}) is achieved for the wav2vec2.0 model of the compressed architecture 11-2560-6bit against the individually re-tuned system. \textbf{2)} A maximum compression ratio of 3.93 times is obtained while incurring no statistically significant WER increase (Sys.~6K vs. Sys.~0 in Tab.~\ref{tab4}). \textbf{3)} Finally, an overall compression and training time speed-up ratio of 2.2 times is achieved (see Training Time in Tab.~\ref{tab4}). In addition, the all-in-one model contains only 94m parameters against 690m in total for 8 individual systems.

\vspace{-0.cm}
\section{Conclusion}
\vspace{-0.1cm}
We propose a novel one-pass multiple ASR systems joint compression and quantization approach using an all-in-one neural model, which enables the simultaneous construction of multiple systems with varying Encoder depths, widths, and quantization precision settings without the need to train and store individual target systems separately. Extensive results demonstrate that all-in-one models produce comparable or even better performance while costing less training time and memory storage against individually trained systems. Future work focuses on building more flexibly layer-wise re-configurable systems.

\newpage
\section{Acknowledgements}
This research is supported by Hong Kong RGC GRF grant No. 14200220, 14200021, Innovation Technology Fund grant No. ITS/218/21.

\bibliographystyle{IEEEtran}
\bibliography{mybib}

\end{document}